\documentclass[slac_one]{revtex4}
\usepackage{epsfig,amsmath,feynarts}
\usepackage{graphicx}
\usepackage{fancyhdr}
\def\pslash#1{{\setbox0=\hbox{$#1$}
  \rlap{\ifdim\wd0>.7em\kern.22\wd0\else\kern.1\wd0\fi /}#1}}
\def\psl{\pslash p}

\def\glui{{\tilde{g}}}
\def\epsbar{{\bar{\epsilon}}}

\def\ghat{{\hat{g}}}
\def\gtilde{{\tilde{g}}}

\begin{document}

\title{{\small{2005 International Linear Collider Workshop - Stanford,
U.S.A.}}\\ 
\vspace{12pt}
Regularization of supersymmetric theories: Recent improvements
} 

%

\author{D.~St\"ockinger}
\affiliation{IPPP, 
University of Durham, Durham DH1 3LE, U.K.}

\begin{abstract}

Recent progress concerning 
regularization of supersymmetric theories is reviewed. 
Dimensional reduction is reformulated in a mathematically
consistent way, and an elegant and general method is presented that
allows to study the supersymmetry-invariance of dimensional reduction
in an easy manner. This method is applied to several supersymmetry
identities at the one- and two-loop level, and thus the extent to
which dimensional reduction is known to preserve supersymmetry is
significantly enlarged.

\end{abstract}

\maketitle

\thispagestyle{fancy}

\section{INTRODUCTION}

Regularization by dimensional reduction (DRED) \cite{Siegel79} is the
most common regularization for supersymmetric theories. In contrast to
ordinary dimensional regularization \cite{HV,BM}, DRED has been shown
to preserve supersymmetry in several cases
\cite{CJN80,BHZ96,STIChecks}. Nevertheless, DRED has always been known
to be mathematically inconsistent \cite{Siegel80}, and as a
consequence there is no general proof that it preserves supersymmetry
in all cases.

The fact that no consistent supersymmetric and gauge-invariant
regularization is known leads immediately to fundamental questions:
Are supersymmetric theories renormalizable at all? Are there genuine
supersymmetry anomalies? These questions have been studied extensively
in a regularization-independent way, and the answers are Yes and No,
respectively \cite{SSTI1,SSTI2,SSTIus}.

But the problems of DRED also lead to very practical questions: (1)
Does the mathematical inconsistency matter in practical calculations?
(2) To what extent is DRED supersymmetric? These questions are
particularly important in view of the SPA (``Supersymmetry Parameter
Analysis'') project \cite{SPA}, where loop calculations within
supersymmetry are required and supersymmetry parameters are defined in
the $\overline{DR}$ scheme. The $\overline{DR}$ scheme is equivalent
to using DRED as a regularization and to perform minimal subtraction
of the divergent terms.

If DRED would break supersymmetry in a certain calculation, additional
(often finite) counterterms would have to be found and added in order
to restore supersymmetry. Hence, in such a case the $\overline{DR}$
scheme as such could not be used and would have to be
modified. Moreover, the technical determination of such
supersymmetry-restoring counterterms is often tedious
\cite{STIChecks}. 

In the present paper we review the results of \cite{DREDPaper}, where
DRED was studied with three aims:
\begin{description}
\item{(1)} DRED should be redefined without a mathematical
  inconsistency. 
\item{(2)} A general method should be found to study the
  supersymmetry-invariance of DRED.
\item{(3)} The general method should actually be applied to verify
  that DRED preserves supersymmetry in several non-trivial cases of
  practical interest.
\end{description}
It turns out that the consistent formulation of DRED allows to prove
the quantum action principle, which is a theorem that can be used as
the key ingredient in the study of symmetry-properties of DRED. We
will describe the consistent formulation of DRED in Sec.\ 2, the
quantum action principle and its role in Sec.\ 3; the desired
method and its applications are discussed in Sec.\ 4.

In the remainder of this introduction we mention another problem of
DRED that is important for the SPA project and the question to what
extent the $\overline{DR}$ scheme can be used for hadronic processes.
In \cite{BKNS,NS} an apparent mismatch between the DRED-result
for the process $gg\to t\bar{t}$ and the expectation from
QCD-factorization has been reported. In the case of massless quarks
instead of $t\bar{t}$, the transition from DRED to ordinary
dimensional regularization for the NLO-corrections involves a simple
convolution with the LO cross section. In the case of massive
$t\bar{t}$ in the final state, however, the transition involves
additional terms that do not have the expected factorized
structure. It is an important task to understand this
puzzling result and to reconcile DRED with QCD-factorization
\cite{factorization}.

\section{MATHEMATICAL CONSISTENCY}

In DRED, only momenta and momentum integrals are continued from 4
to $D$ dimensions, while $\gamma$-matrices and gauge fields remain
4-dimensional objects. Accordingly, two types of metric tensors can
appear in the computation of Feynman diagrams: the 4-dimensional
$g^{\mu\nu}$ can appear e.g.\ in the numerator of vector boson
propagators, and the $D$-dimensional $\ghat^{\mu\nu}$ can appear in
the result of a $D$-dimensional integral $\int d^Dp[p^\mu p^\nu
f(p^2)]$. Defining also a $(4-D)$-dimensional metric tensor
$\gtilde^{\mu\nu}$, they satisfy the following relations:
\begin{subequations}
\label{gmunu}
\begin{align}
g^{\mu\nu} &= \ghat^{\mu\nu}+\gtilde^{\mu\nu} &
g^{\mu\nu}g_{\mu\nu} & =4&
\ghat^{\mu\nu}\ghat_{\mu\nu} &=D &
\gtilde^{\mu\nu}\gtilde_{\mu\nu} &=4-D\\
g^{\mu\nu}\ghat_\nu{}^\rho &=\ghat^{\mu\rho} &
g^{\mu\nu}\gtilde_\nu{}^\rho &=\gtilde^{\mu\rho}&
\ghat^{\mu\nu}\gtilde_\nu{}^\rho &=0
\end{align}
\end{subequations}
These relations correspond to a decomposition of the 4-dimensional
space into a $D$-dimensional subspace and an orthogonal
$(4-D)$-dimensional subspace. Using $\ghat^{\mu\nu}$ and
$\gtilde^{\mu\nu}$ as projectors onto these subspaces we can define
$\hat{a}^\mu=\ghat^{\mu\nu}a_\nu$,
$\tilde{a}^\mu=\gtilde^{\mu\nu}a_\nu$ for any 4-dimensional object
$a^\mu$. In particular this is possible for the 4-dimensional
$\epsilon$-tensor, and we can write down the product
\begin{equation}
\hat{\epsilon}^{\mu\nu\rho\sigma}\,
\tilde{\epsilon}_{\alpha\beta\gamma\delta}\,
\hat{\epsilon}_{\mu\nu\rho\sigma}\,
\tilde{\epsilon}^{\alpha\beta\gamma\delta}\,
.
\label{EpsProduct}
\end{equation}
If we now use the 4-dimensional relation
\begin{equation}
\epsilon^{\mu_1\mu_2\mu_3\mu_4}\,
\epsilon^{\nu_1\nu_2\nu_3\nu_4} 
\propto {\rm det}((g^{\mu_i\nu_j}))
\label{EpsProdRel}
\end{equation}
we can evaluate the product (\ref{EpsProduct}) in two ways. If
(\ref{EpsProdRel}) is applied to the first and second factor in
(\ref{EpsProduct}), we obtain zero, but applying (\ref{EpsProdRel}) 
to factors 1--3 yields $D(D-1)(D-2)(D-3)$ and applying it to factors
2--4 yields $\epsilon(\epsilon-1)(\epsilon-2)(\epsilon-3)$, where
$\epsilon=4-D$. Therefore, evaluating (\ref{EpsProduct}) in these two
ways leads to the two results
\begin{equation}
0 = D(D-1)^2(D-2)^2(D-3)^2(D-4).
\label{Inconsistency}
\end{equation}
This is mathematically inconsistent with $D$ taking arbitrary values.
This fundamental inconsistency of DRED was already discovered in Ref.\
\cite{Siegel80}, and it can be rewritten in several ways involving
$\epsilon$-tensors, $\gamma_5$, or only metric tensors (see e.g.\
\cite{JJReview}). 

It is important to note that the inconsistency (\ref{Inconsistency})
is not derived from the relations (\ref{gmunu}) alone but that the
purely 4-dimensional relation (\ref{EpsProdRel}) is necessary as
well. In \cite{DREDPaper} it is shown that the rules (\ref{gmunu})
are in fact completely consistent. Well-defined objects $g^{\mu\nu}$,
$\ghat^{\mu\nu}$, $\gtilde^{\mu\nu}$ are explicitly constructed such
that the relations (\ref{gmunu}) are satisfied. This ensures that any
application of (\ref{gmunu}) alone will never lead to an inconsistent
result such as (\ref{Inconsistency}). The explicit objects constructed
in Ref.\ \cite{DREDPaper}, however, {\em do not satisfy} eq.\
(\ref{EpsProdRel}), which is why the inconsistency is
avoided.\footnote{In particular, the 4-dimensional metric tensor
  $g^{\mu\nu}$ appearing here {\em does not have} the index
  representation $g^{00}=-g^{ii}=1$ for $i=1,2,3$ and $g^{\mu\nu}=0$
  otherwise. $g^{\mu\nu}$, $\ghat^{\mu\nu}$, $\gtilde^{\mu\nu}$ have
  to be more complicated objects.}

Similarly to $g^{\mu\nu}$, $\ghat^{\mu\nu}$, $\gtilde^{\mu\nu}$,
$\gamma$-matrices can be constructed that satisfy 
\begin{align}
\{\gamma^\mu,\gamma^\nu\}&=2g^{\mu\nu},&
\gamma^\mu\gamma_\mu=4,
\label{GammaRel}
\end{align}
but that do not satisfy further 4-dimensional relations like Fierz
relations. For the evaluation of many Feynman diagrams, eqs.\
(\ref{gmunu}), (\ref{GammaRel}) are sufficient. Therefore, for a wide
range of applications, the consistent version of DRED, where only
(\ref{gmunu}) and (\ref{GammaRel}) may be used, does not differ from
the traditional version, where (\ref{EpsProdRel}) or Fierz identities
might be used in addition.

The consistent formulation of DRED has a crucial consequence. Beyond
the practical evaluation of Feynman diagrams, it allows to give a
general proof of the quantum action principle. This will be exploited
in the next section. 

\section{SUPERSYMMETRY OF DRED AND THE QUANTUM ACTION PRINCIPLE}

It is an important open question whether or to what extent DRED
preserves supersymmetry. So far, several supersymmetry identities
between propagators and/or three-point functions have been shown to be
valid in DRED at the one-loop level \cite{CJN80,BHZ96,STIChecks}. 
However, these checks do not even exhaust all cases of
practical interest; e.g. supersymmetry Slavnov-Taylor identities
relating four-point functions and/or two-loop identities have not been
checked. The traditionally used methods are tedious and by using them
it is hard to extend the checks performed in the literature.

Our second aim is therefore to develop a method that simplifies the
study of supersymmetry in DRED. Generally, supersymmetry Ward or
Slavnov-Taylor identities can be written in the form
\begin{equation}
\delta_{\rm SUSY}\langle T\phi_1\ldots\phi_n\rangle^{\rm DRED}
 \stackrel{(?)}{=} 0,
\label{SymId}
\end{equation}
where the (?) indicates that the identity is not necessarily valid in
DRED but  it is our task to verify it. In Refs.\
\cite{CJN80,BHZ96,STIChecks} this verification was done by explicitly
evaluating all Green functions on the left-hand side of (\ref{SymId})
and checking that they all add up to zero. A drastically simpler way
to check (\ref{SymId}) can be based on the quantum action principle,
which relates the left-hand side of (\ref{SymId}) to a simpler Green
function (see Ref.\ \cite{DREDPaper} for more details, a heuristic
explanation using the path integral and the proof in DRED):
\begin{align}
i\,\delta_{\rm SUSY}\langle T\phi_1\ldots\phi_n\rangle^{\rm DRED}
=& \langle T\phi_1\ldots\phi_n\Delta\rangle^{\rm DRED}\ ,&
\mbox{ where }
\Delta&=\int d^Dx \delta_{\rm SUSY}{\cal L}.
\label{QAP}
\end{align}
The right-hand side of the quantum action principle is a single Green
function involving the insertion of the composite operator $\Delta$,
which can be obtained from the supersymmetry variation of the
regularized Lagrangian. The caveat here is that the quantum action
principle itself is a regularization-dependent statement and it is not
obviously valid in DRED. However, using the consistent formulation
presented in the previous section, (\ref{QAP}) can be shown to hold in
DRED \cite{DREDPaper}. The proof turns out to be analogous to the
corresponding proof for dimensional regularization \cite{BM}.

Hence a given supersymmetry Slavnov-Taylor identity can be checked by
simply verifying that $\langle T\phi_1\ldots\phi_n\Delta\rangle^{\rm
  DRED}$ vanishes. In the next section, we will show how easily this
can be done in several non-trivial examples.

Before applying the quantum action principle explicitly to
supersymmetry Slavnov-Taylor identities, it is convenient to introduce
the notation used in Refs.\ \cite{SSTI1,SSTI2,STIChecks} and
in particular for the Slavnov-Taylor identity of the MSSM
\cite{SSTIus}. All Slavnov-Taylor identities of the form (\ref{SymId})
can be combined into a single identity $S(\Gamma^{\rm DRED})=0$, where
$\Gamma^{\rm DRED}$ is the vertex functional of one-particle
irreducible (1PI) Green functions, regularized using DRED, and $S(\cdot)$ is
a bilinear operator. Particular 1PI Green functions are obtained as
$\Gamma_{\phi_1\ldots}=(\delta\Gamma/\delta\phi_1\ldots)|_{\phi_i=0}$,
and identities for particular Green functions
analogous to (\ref{SymId}) can be rederived by taking derivatives like
\begin{equation}
\left.
\frac{\delta^{n+1}S(\Gamma^{\rm DRED})}
{\delta \phi_n\ldots\delta\phi_1\delta\epsilon}
\right|_{\phi_i=0}\stackrel{(?)}{=}0,
\label{STIId}
\end{equation}
where $\epsilon$ denotes the supersymmetry transformation parameter.

The quantum action principle then takes the form
\begin{align}
S(\Gamma^{\rm DRED}) &=
i[S(\Gamma_{\rm cl})]\cdot \Gamma^{\rm DRED},
\label{STIQAP}
\end{align}
where $[\Delta]\cdot \Gamma^{\rm DRED}$ denotes the insertion of an
operator $\Delta$ into the 1PI vertex functions analogous to the
insertion on the right-hand side of eq.\ (\ref{QAP}). $\Gamma_{\rm
  cl}$ denotes the regularized classical action $\int d^Dx {\cal L}$. 

\section{SUPERSYMMETRY OF DRED UP TO THE TWO-LOOP LEVEL}

We are now going to apply the strategy of the previous section to
study several supersymmetry identities in DRED. That is, we consider
identities of the form (\ref{STIId}) and replace the left-hand side by 
\begin{equation}
\left(i[S(\Gamma_{\rm cl})]\cdot \Gamma^{\rm DRED}
\right)_{\phi_n\ldots\phi_1\epsilon},
\label{STIViolation}
\end{equation}
the 1PI Green function with insertion of the
operator $S(\Gamma_{\rm cl})$ and external fields
$\epsilon\phi_1\ldots\phi_n$. The corresponding
identity (\ref{STIId}) is valid in DRED precisely
if (\ref{STIViolation}) vanishes; in general, (\ref{STIViolation})
constitutes a possible violation of eq.\ (\ref{STIId}).

\subsection{Insertion operator and its Feynman rule}

As a first step, the insertion operator $[S(\Gamma_{\rm cl})]$ has to
be evaluated; it is a major advantage that this operator is universal
and the evaluation has to be performed only once. The result for a
general supersymmetric gauge theory has been given in
\cite{DREDPaper}. We quote here only the part related to matter field
interactions: 
\begin{align}
\label{DeltaRes}
S(\Gamma_{\rm cl})&=
-g\int d^Dx\left[
2(\overline{\psi} P_R\epsilon)(\overline{\glui}P_L\psi)
+2(\epsbar P_L\psi_j)(\overline{\psi}_i P_R\glui_{ij})
+(\overline{\psi}_i\gamma^\mu P_L\psi_j)(\epsbar\gamma_\mu \glui_{ij})
+\ldots\right].
\end{align}
Here $\phi$ and $\psi$ denote the scalar and fermionic components of
chiral multiplets; $\glui$ denotes the gaugino field.
All terms in $S(\Gamma_{\rm cl})$ are four-fermion operators.
In strictly 4 dimensions, where Fierz identities can be used,
$S(\Gamma_{\rm cl})=0$, but in DRED, where Fierz identities are
invalid, the insertion of $S(\Gamma_{\rm cl})$ into Green functions
could lead to non-vanishing breakings (\ref{STIViolation}).


As a second step, the Feynman rule corresponding to the insertion of
$S(\Gamma_{\rm cl})$ into a diagram has to be determined. For our
applications,  diagrams of the basic topology shown in Fig.\
\ref{fig:viol}(a) are most relevant. In these diagrams, the
$\tilde{g}$ and $\overline\psi$ leaving the vertex corresponding to
$S(\Gamma_{\rm cl})$  are connected to a closed fermion loop via
emission of a scalar field $\phi^\dagger$. Additional boson lines can
be attached in the actual applications. 
Denoting the string of $\gamma$-matrices
attached to the external $\psi$ line as $A$ and the
$\gamma$-string associated to the closed fermion loop as $B$, the
Feynman rule for this diagram reads
\begin{align}
\label{ResC}
(\gamma^\mu B \gamma_\mu -2 P_R B^C )P_L A
- 2 P_L A {\rm Tr}(P_R B).
\end{align}
Here $P_{R,L}=\frac12(1\pm\gamma_5)$ and $B^C$ is derived from $B$
using the rule $(\gamma^{\mu_1}\ldots\gamma^{\mu_n})^C=(-1)^n \gamma^{\mu_n}\ldots\gamma^{\mu_1}$.
Again this expression (\ref{ResC}) vanishes identically in strictly four
dimensions. 

In DRED, however, where only the rules (\ref{gmunu}),
(\ref{GammaRel}) can be applied, eq.\ (\ref{ResC}) does not vanish in
general, but it does vanish if $B$ does not contain more 
than three $\gamma$-matrices.
This observation turns out to be  sufficient for all
applications described below.

\subsection{Examples}

The first example we consider is the identity $\frac{\delta^3
  S(\Gamma)}{\delta\phi^\dagger\delta\psi\delta\epsbar} =0$.
Its explicit form reads
\begin{align}
\label{STI1}
0 &=
\Gamma_{\psi \epsbar {Y_{\phi}}_i}\Gamma_{\phi^\dagger\phi_i}-
\Gamma_{\phi^\dagger {Y_{\psi}}_i\epsbar}\Gamma_{\psi\overline{\psi}_i},
\end{align}
and it expresses the fundamental supersymmetry relation between the
$\phi$ and $\psi$ self energies, including the equality of the $\phi$
and $\psi$ masses. Eq.\ (\ref{STI1}) has already been studied
extensively, and it has been shown to be valid in DRED at the one-loop
level in various supersymmetric models \cite{STIChecks}. The checks
performed in Refs.\ \cite{STIChecks} involve the evaluation of all
four Green functions in eq.\ (\ref{STI1}). In particular the necessity
to evaluate also the Green functions involving ${Y_{\phi}}_i$ and
${Y_{\psi}}_i$, corresponding to loop-corrected supersymmetry
transformations of $\phi_i$ and $\psi_i$, makes the checks rather
tedious.

\begin{figure}[ht]
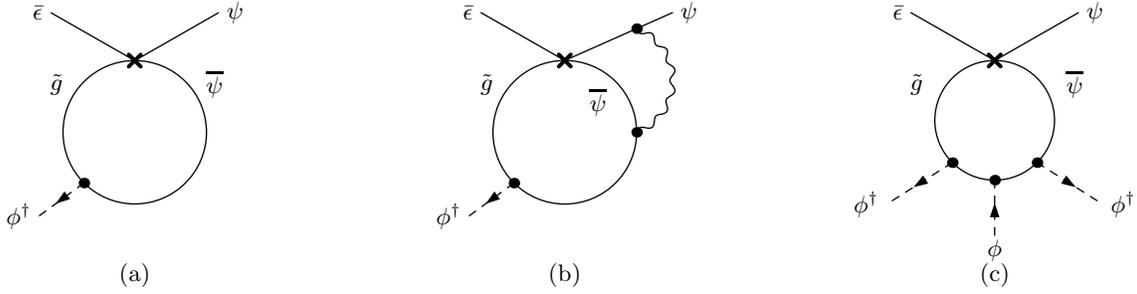

\begin{center}
\unitlength=1.cm%
\begin{feynartspicture}(3.5,4.7)(1,1)
\FADiagram{}
\FALabel(10,-1)[t]{(a)}
\FAVert(10,16){1}
\FALabel(3,20)[r]{$\epsbar$\ }
\FAProp(17,20)(10,16)(0.,){/Straight}{0}
\FALabel(17,20)[l]{\ ${\psi}$}
\FAProp(3,20)(10,16)(0.,){/Straight}{0}
\FALabel(4,14)[r]{$\glui$}
\FALabel(16,14)[l]{$\overline{\psi}$}
\FAProp(10,16)(10,4)(1.,){/Straight}{0}
\FAProp(10,16)(10,4)(-1.,){/Straight}{0}
\FAVert(5.75,5.75){0}
\FALabel(2,3)[r]{${\phi^\dagger}$\ }
\FAProp(5.75,5.75)(2,3)(0.,){/ScalarDash}{1}
\end{feynartspicture}
\hspace{2cm}
\unitlength=1.cm%
\begin{feynartspicture}(3.5,4.7)(1,1)
\FADiagram{}
\FALabel(10,-1)[t]{(b)}
\FAVert(10,16){1}
\FALabel(3,20)[r]{$\epsbar$\ }
\FAProp(19,20)(10,16)(0.,){/Straight}{0}
\FALabel(19,20)[l]{\ ${\psi}$}
\FAProp(3,20)(10,16)(0.,){/Straight}{0}
\FALabel(4,14)[r]{$\glui$}
\FALabel(13.5,12.5)[r]{$\overline{\psi}$}
\FAProp(10,16)(10,4)(1.,){/Straight}{0}
\FAProp(10,16)(10,4)(-1.,){/Straight}{0}
\FAVert(5.75,5.75){0}
\FALabel(2,3)[r]{${\phi^\dagger}$\ }
\FAProp(5.75,5.75)(2,3)(0.,){/ScalarDash}{1}
\FAVert(16,18.66){0}
\FAVert(16,10){0}
\FAProp(16,10)(16,18.66)(.7,){/Sine}{0}
\end{feynartspicture}
\hspace{2cm}
\unitlength=1.cm%
\begin{feynartspicture}(3.5,4.7)(1,1)
\FADiagram{}
\FALabel(10,-1)[t]{(c)}
\FAVert(10,16){1}
\FALabel(3,20)[r]{$\epsbar$\ }
\FAProp(17,20)(10,16)(0.,){/Straight}{0}
\FALabel(17,20)[l]{\ ${\psi}$}
\FAProp(3,20)(10,16)(0.,){/Straight}{0}
\FALabel(4,14)[r]{$\glui$}
\FALabel(16,14)[l]{$\overline{\psi}$}
\FAProp(10,16)(10,6)(1.,){/Straight}{0}
\FAProp(10,16)(10,6)(-1.,){/Straight}{0}
\FAVert(6.46,7.46){0}
\FALabel(1,4)[r]{${\phi^\dagger}$\ }
\FAProp(6.46,7.46)(1,4)(0.,){/ScalarDash}{1}
\FAVert(13.54,7.46){0}
\FALabel(19,4)[l]{\ ${\phi^\dagger}$}
\FAProp(13.54,7.46)(19,4)(0.,){/ScalarDash}{1}
\FAVert(10,6){0}
\FALabel(10,1)[t]{${\phi}$}
\FAProp(10,6)(10,1)(0.,){/ScalarDash}{-1}
\end{feynartspicture}
\end{center}
\vspace{-1.cm}
\caption{(a): Basic topology of a diagram involving an insertion of
  $S(\Gamma_{\rm cl})$, eq.\ (\ref{ResC}). Additional boson lines have
  to be attached in the actual 
  diagrams. In the text, the $\gamma$-string attached to the second
  external $\psi$ line is denoted as $A$, the $\gamma$-string
  attached to the closed fermion loop as $B$.
\\(b): A two-loop diagram corresponding to $([S(\Gamma_{\rm
    cl})]\cdot\Gamma^{\rm DRED})_{\phi^\dagger\psi\epsbar}$, eq.\
    (\ref{STI1Viol}).  Such
    two-loop diagrams involving a virtual vector boson are the only
    ones where the $\gamma$-string $B$ can contain three
    $\gamma$-matrices after 
  integration over the fermion loop momentum.
\\(c):One-loop diagram corresponding to $([S(\Gamma_{\rm
      cl})]\cdot\Gamma^{\rm DRED})_{ 
\phi^\dagger\phi^\dagger
\phi\psi\epsbar}$, eq.\ (\ref{phi4viol}).}
\label{fig:viol}
\end{figure}

Examining identity (\ref{STI1}) becomes almost trivial if the quantum
action principle is used. The possible violation of (\ref{STI1}) is
given by 
\begin{equation}
\label{STI1Viol}
\left(i[S(\Gamma_{\rm cl})]\cdot \Gamma^{\rm DRED}
\right)_{\phi^\dagger\psi\epsbar},
\end{equation}
and the diagram in Fig.\ \ref{fig:viol}(a) is the single one-loop
contributing to this violation. In this diagram, the $\gamma$-string
$B$, corresponding to the closed loop, contains at most two
$\gamma$-matrices. Hence, the expression (\ref{ResC}) and the
whole diagram vanish, and thus there is no violation of (\ref{STI1})
in DRED at the one-loop level.

It is possible to extend this analysis to the two-loop level. There
are several two-loop diagrams corresponding to the possible violation
(\ref{STI1Viol}); the one with the most $\gamma$-matrices in the
fermion loop is shown in Fig.\ \ref{fig:viol}(b). After integrating over
the fermion loop momentum, this diagram contains
only up to three $\gamma$-matrices in the $\gamma$-string $B$, and
therefore it vanishes. It can be easily seen that the same is true for
all two-loop diagrams contributing to (\ref{STI1Viol}). This shows
that the propagator identity (\ref{STI1}) is valid in DRED even at
the two-loop level.

In a similar way one can derive and study an identity relating the
loop-corrected supersymmetry transformations $\Gamma_{\psi \epsbar
  {Y_{\phi}}_i}$ and $\Gamma_{\phi^\dagger {Y_{\psi}}_i\epsbar}$. Such
identities are generally important because they express the fact that
in spite of loop corrections to the supersymmetry transformations, the
supersymmetry algebra still holds. In turn, this is a necessary
condition for identities like (\ref{STI1}) to correspond to
supersymmetry relations. In Refs.\ \cite{STIChecks} several
supersymmetry algebra identities have been discussed and verified at
the one-loop level. 

We can use again the quantum action principle to write the possible
violation of the identity relating $\Gamma_{\psi \epsbar
  {Y_{\phi}}_i}$ and $\Gamma_{\phi^\dagger {Y_{\psi}}_i\epsbar}$ in
the form (\ref{STIViolation}). It turns out that there is no
corresponding one-loop diagram at all. Moreover, it can be easily
shown that all two-loop diagrams contributing to the violation of the
identity vanish \cite{DREDPaper}. 

Therefore, in this approach not only the results found in
\cite{STIChecks} become completely obvious, but it is also very easy
to extend the results to the two-loop level.

The final example we consider concerns the $\phi^4$
interaction. It is well-known that in supersymmetric models the
$\phi^4$ terms in the scalar potential are completely determined in
terms of gauge and Yukawa couplings and do not involve free
parameters. This is in particular the origin of the Higgs boson mass
predictions in the MSSM, which has been computed up to the two-loop
level (see \cite{HHW} for a review). However, the corresponding
Slavnov-Taylor identity describing the correct treatment of the
$\phi^4$ interaction at the loop level has never been verified, not
even at the one-loop level.

This Slavnov-Taylor identity for the $\phi^4$ interaction is given by
$\frac{\delta^5 S(\Gamma)}
{\delta\phi^\dagger\delta\phi\delta\phi^\dagger
\delta\psi\delta\epsbar}=0$, and according to (\ref{STIViolation}) its
possible violation is given by 
\begin{equation}
\left(i[S(\Gamma_{\rm cl})]\cdot \Gamma^{\rm DRED}
\right)_{\phi^\dagger\phi\phi^\dagger
\psi\epsbar}.
\label{phi4viol}
\end{equation}
The diagram in Fig.\ \ref{fig:viol}(c) is the only one-loop diagram
contributing to this Green function (up to permutations). As in the
previous cases, after integrating over the fermion loop momentum, the
$\gamma$-string $B$ can contain at most three $\gamma$-matrices,
here corresponding to $\psl_i$ for the three independent incoming
momenta $p_i$. Hence the violation (\ref{phi4viol}) vanishes and the
$\phi^4$ Slavnov-Taylor identity is valid in DRED at the one-loop
level.

\section{CONCLUSIONS}

We have studied DRED with three aims presented in the
introduction. First DRED could be redefined in a mathematically
consistent way. The difference to the traditional formulation concerns
only the validity of Fierz and similar relations but is not relevant
in a wide range of applications. 

In a second step a general method to
study supersymmetry identities in DRED was developed based on the
quantum action principle (\ref{QAP}), (\ref{STIQAP}). Using the
consistent formulation of DRED, the quantum action principle could be
established in DRED. Supersymmetry Slavnov-Taylor identities of the
form  (\ref{STIId}) are then generally
violated by the expression 
$\left(i[S(\Gamma_{\rm cl})]\cdot \Gamma^{\rm DRED}
\right)_{\phi_n\ldots\phi_1\epsilon}$,
eq.\ (\ref{STIViolation}). This is a Green 
function involving the insertion of the operator $S(\Gamma_{\rm cl})$,
which has been evaluated explicitly, see eq.\ (\ref{DeltaRes}). 

Finally, this method has been applied to study several supersymmetry
identities of practical interest. The identities for propagators, eq.\
(\ref{STI1}), and for the corresponding supersymmetry transformations
have been considered already in Refs.\ \cite{STIChecks}, but only at
the one-loop level. We have shown here that rederiving the one-loop
results using the described method is very easy, and we could present
the verification of these 
identities at the two-loop level. In addition, the identity for the
$\phi^4$ interaction has been verified at the one-loop level.

In conclusion, the  status of DRED has been improved by 
establishing mathematical consistency, the quantum
action principle and the validity of supersymmetry identities up to
the two-loop level. A crucial outcome is that using
the developed method, studying supersymmetry identities is
dramatically simplified beyond the considered examples. For the
future it will be important to further 
study the properties of DRED, in particular to verify
that DRED preserves supersymmetry at least to the level
required for loop calculations of LHC- or ILC-observables. This goal
will require more work, but it has come within reach with the results
presented here. 


\end{document}